\documentclass[12pt,draftclsnofoot,onecolumn]{IEEEtran}
\usepackage{graphicx,cite,epsfig,amssymb,amsmath,subfigure,url,stfloats,latexsym,color,setspace}

\begin{document}
\begin{spacing}{1.58}
\markboth{IEEE Wireless Communications, vol. XX, no. XX, Month 2016.} {Chen, Ng \& Chen:
Secure Wireless Information and Power Transfer \ldots}

\title{Secrecy Wireless Information and Power Transfer: Challenges and Opportunities }

\author{\authorblockN{
Xiaoming~Chen, \IEEEmembership{Senior Member, IEEE}, Derrick~Wing~Kwan~Ng,
\IEEEmembership{Member, IEEE}, and Hsiao-Hwa~Chen,
\IEEEmembership{Fellow, IEEE}
\thanks{Xiaoming~Chen (e-mail: {\tt chenxiaoming@nuaa.edu.cn}) is with the College of Electronic and Information Engineering, Nanjing University of Aeronautics and Astronautics, Nanjing, China. Derrick~Wing~Kwan~Ng (e-mail: {\tt wingn@ece.ubc.ca}) is with the School of Electrical Engineering and Telecommunications, the University of New South Wales, NSW, Australia. Hsiao-Hwa~Chen (e-mail: {\tt hshwchen@ieee.org}) is with the Department of Engineering Science, National Cheng Kung University, Tainan, Taiwan.}
}}

\maketitle

\vspace{-0.75in}
\begin{abstract}
Wireless information and power transfer (WIPT) enables more sustainable and resilient communications owing to the fact that it avoids frequent battery charging and replacement. However, it also suffers from possible information interception due to the open nature of wireless channels. Compared to traditional secure communications, secrecy wireless information and power transfer (SWIPT) carries several distinct characteristics. On one hand, wireless power transfer may increase the vulnerability of eavesdropping, since a power receiver, as a potential eavesdropper, usually has a shorter access distance than an information receiver. On the other hand, wireless power transfer can be exploited to enhance wireless security. This article reviews the security issues in various SWIPT scenarios, with an emphasis on revealing the corresponding challenges and opportunities for implementing SWIPT. Furthermore, we provide a survey on a variety of physical layer security techniques to improve secrecy performance. In particular, we propose to use massive multiple-input multiple-output (MIMO) techniques to enhance power transfer efficiency and secure information transmission simultaneously. Finally, we discuss several potential research directions to further enhance the security in SWIPT systems.
\end{abstract}

\IEEEpeerreviewmaketitle

\vspace{0.35in}
\section{INTRODUCTION}
Since the pioneer work done by Grover and Sahai \cite{WIPT1}, wireless information and power transfer (WIPT) has spurred considerable interests from academia and industry, especially in the area of wireless communications. Nowadays, most wireless devices are powered via power cables or battery replacement, which limits the scalability, sustainability, and mobility of wireless communications. In practice, wireline charging and battery replacement may be infeasible or incur a high cost under some conditions. For instance, it is impossible to replace the battery of implanted medical devices in human bodies. Besides, wireline charging and battery renewal shortens working period of wireless mobile devices. As a result, radio frequency (RF) signal based WIPT was proposed as a complementary technology to prolong the lifetime of power-limited nodes or networks in a relatively simple and reliable way.

In WIPT systems, information and power signals are carried by the same RF-wave. Then, information is recovered at information receivers, and electromagnetic energy is harvested and converted into electric energy at power receivers \cite{WIPT2}. However, due to the open nature of wireless channels, information signals are also received by power receivers or other unintended receivers, resulting in potential information leakage. Traditionally, upper-layer encryption techniques are utilized to guarantee secure information transmission. However, conventionally cryptography techniques may consume a large amount of energy due to associated high computational complexity. It may lead to a low system energy efficiency and become a high burden on power transfer to support energy for encryption and decryption. Recently, physical layer security (PHY-security) was proved to be an effective alternative to provide secure communications by exploiting the characteristics of wireless channels, such as fading, noise, and interferences \cite{PHY-security}. Especially in WIPT systems, a power signal, dedicated for transferring wireless power, can be exploited to confuse the eavesdroppers, so as to enhance wireless security. Meanwhile, the information signal can also act as a power source to increase the amount of harvested energy at the power receivers. Thus, PHY-security techniques can be naturally applied to secrecy WIPT (SWIPT).

From an information-theoretic viewpoint, the essence of PHY-security is to maximize the secrecy rate, which is defined as a rate difference between the main channel from the transmitter to the legitimate receiver, and the wiretap channel from the transmitter to the eavesdropper \cite{SC}. Hence, it is necessary to enhance the signal received at the legitimate information receiver and to impair the signal received at the eavesdroppers simultaneously. However, the design of PHY-security techniques is a non-trivial task in SWIPT systems, since it also needs to achieve another important objective, namely the wireless power transfer efficiency. In other words, the design of SWIPT systems can be naturally formulated as a dual-objective optimization problem, one for secure information transmission, the other for efficient power transfer. In general, these two objectives may conflict with each other. Let us take a look at a simple example. If a power receiver is a potential eavesdropper, then the effort in improving the power transfer efficiency via increasing the power of information signal may result in a loss of secrecy rate. Thus, it is imperative to strike a good balance between information transmission security and power transfer efficiency. In \cite{SWIPT1}, the tradeoff between information transmission security and power transfer efficiency is formulated into two different problems: the first problem maximizes the secrecy rate subject to individual minimum harvested energy requirement, while the second problem maximizes the weighted sum of harvested energy subject to a minimum required secrecy rate constraint. The two problems are solved jointly by designing spatial beamformers for information and power signals at a multiple-antenna base station. It is generally known that channel state information (CSI) at a transmitter has a great impact on the performance of multiple-antenna systems. However, the CSI may be imperfect due to channel estimation errors or limited CSI feedback. Thus, the authors in \cite{SWIPT2} designed a robust beamforming scheme for multiple-antenna SWIPT systems in the presence of channel uncertainty of information and power receivers. To further improve the performance of SWIPT, the authors in \cite{SWIPT3} proposed to use both artificial noise (AN) and power signal to confuse eavesdroppers and increase the amount of harvested energy simultaneously through proper spatial beamforming. Moreover, SWIPT in relaying systems \cite{SWIPT4}, multicarrier systems \cite{SWIPT5}, and cognitive radio networks \cite{SWIPT6} were also investigated, respectively.

In general, it is impossible to find a universal solution for guaranteed performance of SWIPT, since there are a variety of scenarios with different information transmission, power transfer, and signal eavesdropping schemes. Moreover, for enabling PHY-security, there are a lot of viable techniques. Thus, it is necessary to adopt different PHY-security techniques according to different systems. In this article, we first investigate the security issues in various WIPT scenarios, and point out the fundamental challenges for enabling communication security. Then, we provide a survey of several effective physical-layer techniques to realize SWIPT, with an emphasis on revealing their performance advantages and limitations. Furthermore, we propose to use massive MIMO techniques to enhance information transmission security and to improve power transfer efficiency simultaneously. Finally, we discuss some future research directions.


\vspace{0.25in}
\section{CHALLENGING ISSUES IN SWIPT}
Security is a common problem in wireless communication networks due to the broadcast nature of wireless channels. To be more specific, the information sent to a legitimate receiver is also received by unintended receivers, namely eavesdroppers. From the perspective of PHY-security, in order to guarantee communication security, it is necessary to enhance the signal received at the legitimate receiver and to impair the signal seen at the eavesdroppers simultaneously. However, in SWIPT systems, information transmission security and power transfer efficiency are equally important. The dual system objectives introduce a paradigm shift and bring in new challenging issues to the design of PHY-security in SWIPT systems. In particular, information signals and power signals may compete with each other for the limited system resources. Besides, the objectives of secure information transmission and efficient power transfer may not align or even conflict with each other. In what follows, we give an overview of security issues in different SWIPT systems.

\subsection{SWIPT in Broadcasting Channels}

\begin{figure}[h]
\centering \vspace{0.1in}
\includegraphics [width=0.65\textwidth] {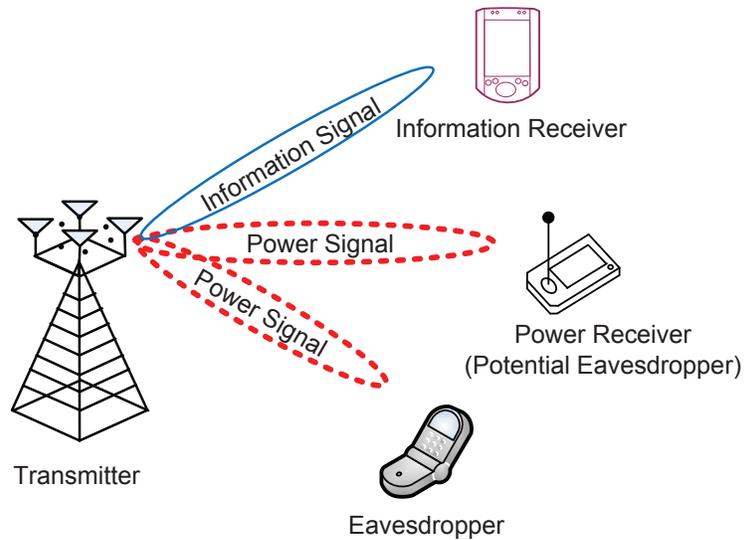}
\caption {A SWIPT scenario in broadcasting channels.} \label{Fig1}
\end{figure}

In SWIPT scenario, a central node plays the role as both an information and a power transmitter, which broadcasts information and power signals to the corresponding receivers, as shown in Fig. \ref{Fig1}. However, the power receivers also receive the information signal. If the power receivers are malicious, eavesdropping may take place \cite{BC}. Thus, they should be treated as potential eavesdroppers for guaranteeing SWIPT. The challenges of SWIPT based on PHY-security in such a scenario are mainly three-fold:
\begin{enumerate}
\item Short distance interception: In practice, information and power receiving circuits have very different power sensitivities. Typically, the minimum requirement for received power at an information receiver is -60 dBm, while that of a power receiver is -10 dBm \cite{SWIPT1}. In order to satisfy the sensitivity requirements, a power receiver is usually located closer to the transmitter than an information receiver. In other words, the power receiver, acting as an potential eavesdropper, may have a shorter signal propagation distance than the legitimate information receiver. As a well known fact, the receive signal quality is a decreasing function of propagation distance. Thus, the signal received at a power receiver (potential eavesdropper) may be much stronger than that received at the legitimate receiver, resulting in a high risk of information interception.
\item Cooperative eavesdropping: In SWIPT, a power receiver may be a potential eavesdropper, and thus it is unlikely to impair the signal at the eavesdropper as much as possible in order to fulfill the requirement of energy harvesting. In this context, malicious power receivers may eavesdrop information cooperatively. For instance, they might cooperate with each other to perform joint signal detection. Thus, the quality of intercepted signal may be much better than that of the signal received at the information receiver, leading to potential information leakage.
\item Inter-user interference: The transmitter broadcasts multiple information and power signals simultaneously. Then, the information receiver suffers strong interference from not only undesired information signals, but also the power signals. However, for conventional interference mitigation schemes, mitigating inter-user interference may lead to a weak RF signal at the power receiver, which are not applicable to the SWIPT in broadcasting channels.
\end{enumerate}

Although SWIPT in broadcasting channels faces several challenging issues, there are some opportunities to realize secure transmission. For example, the transmitter may acquire the channel state information (CSI) of the power receivers (potential eavesdroppers), since they are the legitimate users for harvesting power of the SWIPT systems. Then, the transmitter can design suitable transmission schemes according to the instantaneous CSI, so as to optimize the secrecy performance while satisfying the QoS requirements of energy harvesting at the power receivers.

\subsection{SWIPT in Relaying Systems}

\begin{figure}[h]
\centering \vspace{0.1in}\
\includegraphics [width=0.65\textwidth] {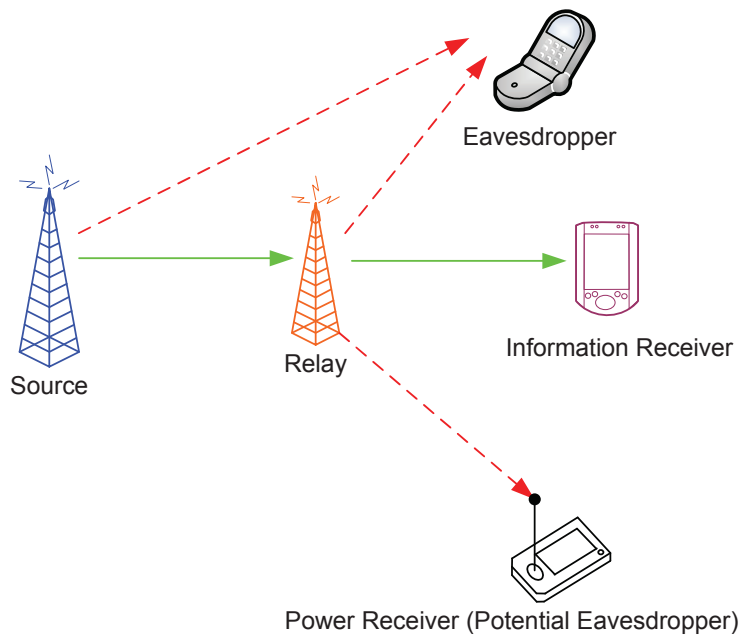}
\caption {A SWIPT scenario in relaying networks.} \label{Fig2}
\end{figure}

Cooperative relaying can shorten transmission distance and provide a diversity gain, and thus it is a commonly used performance-enhancing technique in conventional wireless communication networks. Similarly, cooperative relaying can also be adopted to improve the performance of SWIPT, as shown in Fig. \ref{Fig2}. In SWIPT systems, there are two fundamental relaying modes. The first mode is wireless powered relaying communication, where a relay without self power supply splits up the received signal sent by the source into two components, one for information relaying, the other for energy harvesting. Then, the relay forwards the information signal to the destination with harvested energy. The second mode is that the relay with self power supply forwards the information signal from the source received previously to the information receiver and sends the power signal to the energy receiver concurrently. For both relaying modes, there are some common challenging issues to ensure security in WIPT networks as follows.
\begin{enumerate}
\item Untrusted relay: In a untrusted relay model, although the relay is a cooperative node, it may be a potential eavesdropper or a malicious node \cite{PHY-security}. In the SWIPT, a untrusted relay may increase interception probability significantly. For instance, if the relay is powered by received signal from the source, it may legally receive the signal and decode it. Even if the relay only forwards the signal without energy harvesting, it may pollute the information signal with a power signal intentionally, and thus weaken the signal quality at the information receiver.
\item Vulnerable transmission: Cooperative relaying transmission usually requires two orthogonal time slots to complete information transmission. Thus, an external eavesdropper can receive two copies of the information signal, which can be combined to improve the signal-to-noise-ratio (SNR). Especially in SWIPT, the power receiver, as a potential eavesdropper, may receive two strong copies of signals in order to satisfy the requirement on receiver sensitivity. Thus, the signal quality at the eavesdropper might be better than that at the information receiver. In other words, the transmission is susceptible to eavesdropping.
\end{enumerate}

Nevertheless, there are still some powerful relaying techniques to enhance both information transmission security and power transfer efficiency. In particular, it is possible to perform multiple-relay cooperative transmission. The relays can cooperative with each other to create a virtual multiple-input multiple-output (MIMO). For example, some relays can share their antennas to perform information beamforming to a legitimate information receiver, while the others can adopt power beamforming to transfer wireless power to the power receivers.

\subsection{SWIPT in Interference Networks}

\begin{figure}[h]
\centering \vspace{0.1in}
\includegraphics [width=0.655\textwidth] {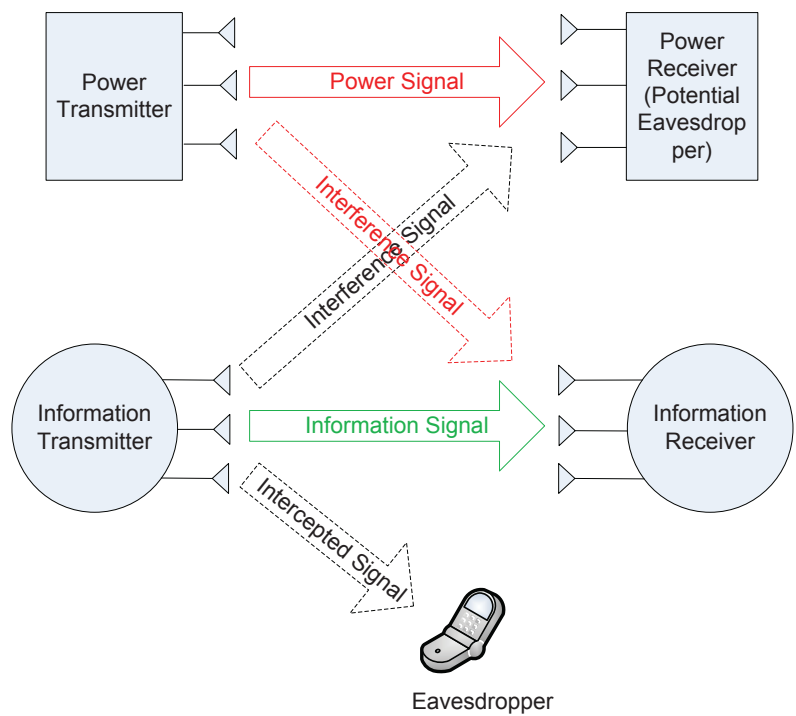}
\caption {A SWIPT scenario in interference networks.} \label{Fig3}
\end{figure}

In an interference network, multiple information and power transceivers communicate with each other over the same channel, as shown in Fig. \ref{Fig3}. For a power receiver, it can receive multiple signal streams from all the transmitters, which can increase the amount of harvested energy. However, the concurrent information and power transmissions will generate severe co-channel interferences at an information receiver, which results in a low received signal-to-interference-plus-noise ratio (SINR) and a high risk of information leakage. Next, we discuss the following challenging issues for SWIPT in interference networks.
\begin{enumerate}
\item Uncoordinated transmission: To simultaneously enhance information transmission security and improve power transfer efficiency, it is better to coordinate the transmissions between information and power transmitters. However, information and power transmitters are geographically separated in heterogeneous networks, and it is difficult for them to exchange information and coordinate transmissions between each other.
\item Unavailability of CSI: According to the theory of wireless power transfer, power receivers convert the harvested RF signals into electric energy. In other words, the power receivers are not necessarily equipped with baseband circuits for signal processing. Then, the power receivers may not be able to perform channel estimation to get the CSI. As a result, the transmitters can not adaptively choose secure schemes according to instantaneous CSI, since the CSI is usually fed back from the receivers.
\item Conflicting objectives: For SWIPT in interference networks, interference has a detrimental impact on information transmission, but is beneficial for power transfer. It may reduce the amount of harvested energy at a power receiver, while mitigating the interference at an information receiver for improving the secrecy rate. It is a challenging task to balance the two conflicting objectives for the design of SWIPT in interference networks.
\end{enumerate}

Note that the co-channel interference in interference networks can also be used to enhance information transmission security and power transfer efficiency, if the transmission scheme is properly designed. First, the multiple signal streams can be exploited to increase the amount of harvested energy at the power receivers. Second, the undesired signals can act as artificial noise to confuse the eavesdroppers to improve the secrecy performance.

\subsection{SWIPT in Wireless Powered Communication Networks}

\begin{figure}[h]
\centering \vspace{0.1in}
\includegraphics [width=0.765\textwidth] {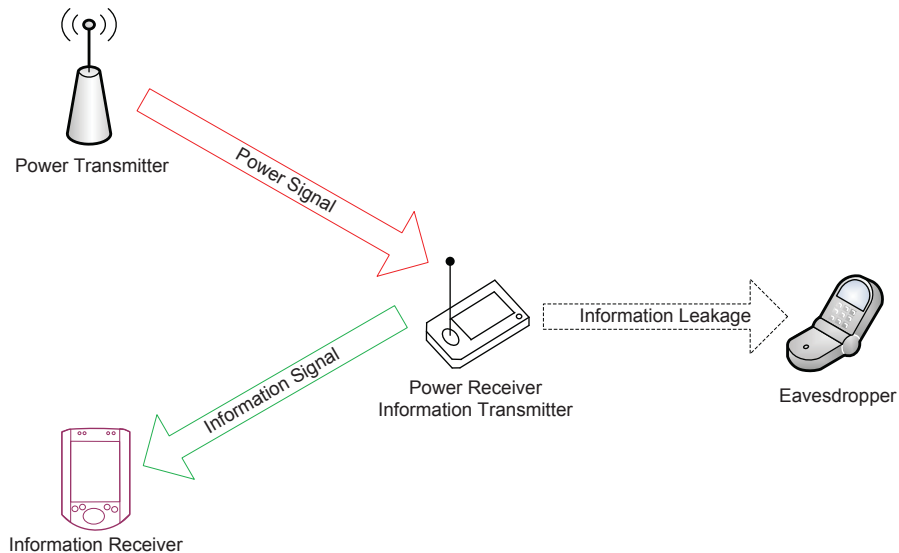}
\caption {A SWIPT scenario in wireless powered communication networks.} \label{Fig4}
\end{figure}

In wireless powered communication networks, a power receiver uses harvested power from RF signals to send messages to an information receiver, as shown in Fig. \ref{Fig4}. A practical application for such a network is medical care. Specifically, the implanted medical devices transmit the information to the instrument outside with the harvested energy. With respect to other SWIPT scenarios, SWIPT in wireless powered communication networks has some special challenging issues, listed as follows.
\begin{enumerate}
\item Difficulty in performing joint resource allocation: Wireless powered communication combines information and power transfer more closely than general WIPT. In particular, the harvested power at a power receiver (also as an information transmitter) affects the performance of information transmission directly. For PHY-security, the secrecy rate is not an increasing function of transmit power. Thus, it is necessary to allocate the resources between power transfer and information transmission to optimize the secrecy performance. For example, a time slot should be partitioned into two orthogonal sub-time slots. The first sub-time slot is used for power transfer and the second is exploited for information transmission. In fact, the optimal partition of a time slot is coupled with transmit power at the power transmitter. Thus, it makes sense to allocate these resources jointly. Yet, the resulting optimization problem is generally non-convex, which does not facilitate the design of computational efficient resource allocation algorithms.
\item Weak anti-eavesdropper capability: For SWIPT in wireless powered communication networks, the transmit power at an information transmitter is obtained through RF energy harvesting, and thus there is only a limited available power. In this case, it is unlikely to adopt sophisticated anti-eavesdropper schemes at the information transmitter, resulting in a weak anti-eavesdropping capability. As a simple example, if there is no enough power, the power of artificial noise is too low to interfere with the eavesdropper effectively.
\end{enumerate}

For enabling SWIPT in wireless powered communication networks, it is better to give the task of resource allocation and anti-eavesdropping to the power transmitter, which may have enough power to support and facilitate sophisticated secure schemes to guarantee SWIPT.

\subsection{SWIPT in Cognitive Radio Networks}

\begin{figure}[h]
\centering \vspace{0.1in}
\includegraphics [width=0.765\textwidth] {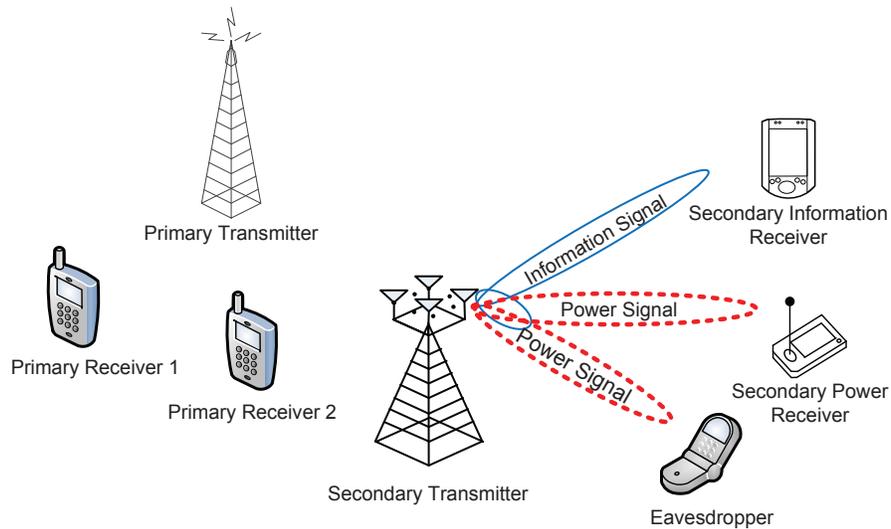}
\caption {A SWIPT scenario in cognitive radio networks.} \label{Fig5}
\end{figure}

Recently, SWIPT in cognitive radio networks (CRNs) has received a lot of attentions. In this case, a secondary transmitter broadcasts both information and power signals over a licensed spectrum band owned by a primary network, as shown in Fig. \ref{Fig5}. In general, the precondition for the secondary network to perform WIPT in licensed spectrum is that the activities in the secondary network will not degrade the QoS of the primary network. However, to enable SWIPT in such networks, we need to tackle the following challenging issues:
\begin{enumerate}
\item Open architecture: Due to the open and dynamic nature of cognitive radio architecture, various unknown information and power devices are allowed to opportunistically access the licensed spectrum. This is vulnerable to eavesdropping as a power receiver, as a potential eavesdropper might obtain more knowledge of information transmitter due to the signal exchanges during cooperative spectrum sensing.
\item Restricted secure scheme: In order to fulfil the precondition for spectrum access, there are limited degrees of freedoms available for information and power transfer, resulting in a performance degradation. For instance, the transmit direction and power for artificial noise is further limited due to interference constraint.
\item Interference management: Primary and secondary networks coexist over the same spectrum. Thus, secondary information receivers may encounter the interferences from the primary network, resulting in a low quality of received signal and a low secrecy rate.
\end{enumerate}


SWIPT in CR networks opens up new opportunities for cooperative communications between the primary and secondary systems at both the information and power harvesting levels. In particular, the secondary transmitter can transmit both secret information and power signals to the secondary receivers, while it charges energy limited primary receivers wirelessly, in exchange of utilizing the licensed spectrum. This approach provides more incentives for both systems to cooperate and therefore improves the overall system performance.

\vspace{0.25in}
\section{PHY-SECURITY TECHNIQUES FOR SWIPT}

As mentioned earlier, SWIPT faces a variety of challenging issues. Thus, it is necessary to adopt some effective PHY-security techniques to enhance information security. However, different from traditional PHY-security techniques in secure communications, the techniques designed for SWIPT networks focus not only on information transmission security, but also on power transfer efficiency. In the sequel, we introduce several powerful PHY-security techniques for guaranteeing SWIPT.

\subsection{Multiple Antenna Techniques}
Multiple antenna technique is commonly used in various SWIPT scenarios to improve communication security \cite{Multi-Antenna}. By exploiting the spatial degrees of freedom offered by multiple antennas, it is possible to enhance the received signals at both information receivers and power receivers to weaken the received signal at eavesdroppers simultaneously, so as to improve the secrecy performance and to increase the amount of harvested energy. For instance, if information signal is transmitted in the null space of the eavesdroppers' channels, the eavesdroppers can not overhear any information. Also, the power signal can be used to confuse the eavesdroppers, while still meeting the requirement of energy harvesting at the power receivers via adaptively adjusting the transmit beamforming.

\subsection{Artificial Noise}
From an information-theoretic viewpoint, the performance of PHY-security is determined by the rate difference between the main channel from the transmitter to the legitimate receiver, and the wiretap channel from the transmitter to the eavesdropper. Thus, if we can impair the intercepted signal, while causing minimal interference to the signal received at the legitimate receiver, it is likely to improve the secrecy performance. Inspired by this idea, artificial noise is introduced into SWIPT networks \cite{AN}. In particular, the power signal can be used as artificial noise, and thus there is no need to consume extra power for a dedicated artificial noise. Moreover, it is likely to send the artificial noise from a friend jammer. Especially in SWIPT, the jammer can first harvest energy from the power transmitter, and then transmits the jamming signal with the harvested energy. Thus, the jammer does not need self power supply.

The key in designing an artificial noise is to adjust the transmit direction in order to avoid the interference to an information receiver. For example, if full CSI of the information receiver is available, it is possible to transmit artificial noise in the null space of the main channel. However, if the CSI regarding the information receiver is imperfect, the artificial noise will leak into the main channel. Thus, it is imperative to adjust the transmit direction, so as to achieve a good balance between the interference to the eavesdroppers and that to the information receiver. On the other hand, the dual use of artificial noise can facilitate efficient wireless power transfer and ensure communication security. Specifically, artificial noise is able to degrade the channels between the transmitter and the potential eavesdroppers and acts as an energy source for power harvesting. Thus, the design of artificial noise should leverage the tradeoff between confusing the power receivers when they perform information decoding and increasing the amount of harvested power at the power receivers.

\subsection{Resource Allocation}
In SWIPT communication systems, there are a variety of available resources, i.e., antenna, time, frequency, and power. In practice, resource allocation in SWIPT plays an important role for performance enhancement. However, it is not a straight forward task in performing efficient resource allocation. This is because resource allocation could affect the performance of information receivers, power receivers, and eavesdroppers simultaneously in SWIPT systems. In fact, the coupling between the aforementioned performance metrics complicates the algorithm design. Additionally, the performance of resource allocation depends heavily on the CSI at the transmitter. If there is full CSI available, it is possible to perform optimal resource allocation in global sense. However, if CSI is unavailable, only a fixed resource allocation scheme can be adopted. In most cases, the transmitter can only get partial and imperfect CSI, and thus it is difficult to design an optimal resource allocation scheme to realize both secure information transmission and efficient power transfer simultaneously.

\subsection{Relay Selection}
Path loss is a detrimental effect on both information transmission and power transfer. By applying cooperative relaying techniques, it is possible to shorten the distance of signal propagation, and thus enhance the performance of SWIPT systems. Especially, multiple-relay cooperation can achieve a better performance. This is because the information receiver, the power receiver, and the eavesdropper are geographically separated. Then, these relays can play different roles according to their locations. For instance, the relays close to the information receiver forward the messages from the transmitter, the ones close to the power receiver send the power signals, and the ones close to the eavesdropper transmit the artificial noise. Through relay selection, it is possible to effectively improve the performance of SWIPT. However, cooperative relaying requires information exchanges between multiple relays, leading to a high overhead.

\vspace{0.25in}
\section{ENHANCED SWIPT BY MASSIVE MIMO TECHNIQUES}
As mentioned above, the design of SWIPT consists of two conflicting objectives, including secure information transmission and efficient power transfer. By exploiting conventional physical layer techniques, e.g., multi-antenna technique, artificial noise, resource allocation, and relay selection, the performance gain is limited, especially under some adverse conditions, such as short-distance and cooperative eavesdropping. Thus, more effective techniques are needed to enhance the performance.

\begin{figure}[h]
\centering \vspace{0.1in}
\includegraphics [width=.87\textwidth] {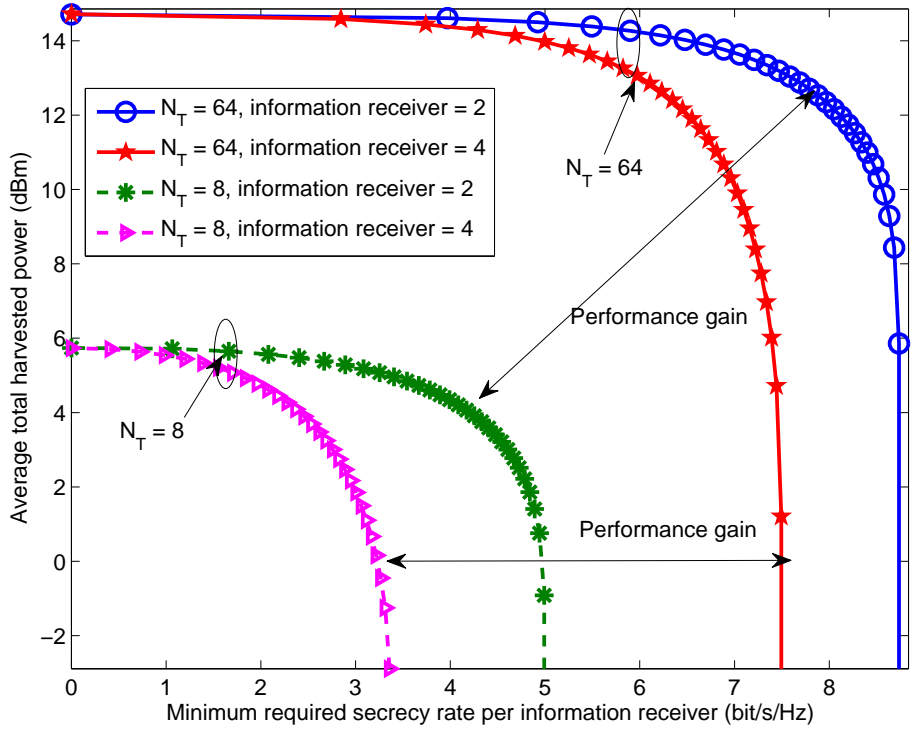}
\caption {The tradeoff region of the average total harvested power (dBm) and the minimum required secrecy rate (bit/s/Hz) per information receiver. The double-sided arrows indicate the performance gain brought in by massive MIMO. The carrier frequency is 915 MHz and the receivers are located 10 meters away from the transmitter. The total transmit power, noise power, transceiver antenna gain, and RF-to-electrical energy conversion loss are set to 36 dBm, -23 dBm, 10 dBi, and 3 dB, respectively. The multipath fading coefficients are modelled as independent and identically distributed Rician random variables with a Rician $K$-factor of 6 dB.} \label{Fig6}
\end{figure}

Recently, massive MIMO techniques \cite{Massive} were introduced to improve the performance of SWIPT significantly. On one hand, by exploiting its large array gain, information and power signal beams can be steered towards the information receivers and power receivers more accurately, respectively. On the other hand, due to high-resolution of spatial beamformers, the information leakage to unintended receivers can be reduced substantially. Theoretically, if the number of antenna is sufficiently large, the leakage information is expected to be negligible. In addition, massive MIMO techniques simplify the associated signal processing for information and power transmission. Even with a low-complexity transmission scheme, e.g. maximum ratio transmission (MRT), it is able to achieve an asymptotically optimal performance. More importantly, due to channel hardening in massive MIMO systems \cite{ChannelHardening}, the performance analysis and optimization become simpler. In what follows, we show the performance gain of using massive MIMO in SWIPT through a simple example. Fig. \ref{Fig6} depicts an example of SWIPT systems in a single cell network with perfect CSI. We show the average total harvested power (dBm) versus the minimum required secrecy rate (bit/s/Hz) per information receiver. In particular, a transmitter equipped with $N_{\mathrm{T}}$ antennas is serving different number of single-antenna information receivers and two multiple-antenna power receivers, each of which is equipped with $N_\mathrm{R}=2$ receive antennas. As can be observed, with an optimal design of information signals and power signals, the trade-off region of the average total harvested power and the minimum required secrecy rate increases significantly with $N_{\mathrm{T}}$, in particular with massive MIMO, i.e., $N_{\mathrm{T}}\geq64$. Besides, the average total harvested power decreases with the number of information receivers, which illustrates the conflicting system design objectives of communication security and total harvested power in SWIPT networks.

\vspace{0.25in}
\section{FUTURE RESEARCH DIRECTIONS}
SWIPT will continue to be a critical research topic and there are a variety of challenging issues as mentioned earlier. In what follows, we list some future research directions on SWIPT networks.

\subsection{CSI Acquisition}
As expected, the CSI at a multiple antennas transmitter has a great impact on the performance of SWIPT. If there is global and perfect CSI, it is likely to design the optimal transmit beamformers to achieve the goals for both secure information transmission and efficient power transfer. However, it is not a trivial task to collect the CSI in SWIPT networks. First, the eavesdropper CSI may be unavailable, since the external eavesdropper is usually passive and well hidden. Second, the CSI of the power receivers is difficult to obtain, because the power receivers may not have a baseband circuit. Third, the CSI corresponding to the information receiver may be imperfect, as it is required to be conveyed from the receiver to the transmitter. In this context, it is necessary to design robust secure beamforming for SWIPT with partial and imperfect CSI \cite{SWIPT2}, \cite{SWIPT3}, \cite{SWIPT6}.

\subsection{Multi-Objective System Design}
Another concern for secure beamforming is the multiple system design objectives. Recently, driven by environmental concerns, energy efficiency (EE) has become an important metric for evaluating the performance of wireless communication systems. However, with SWIPT, the EE of wireless power transfer has the same importance as the EE of secrecy information transmission. As a result, multiple conflicting system design objectives arise naturally in system design process, and the application of the solutions for single-objective optimization to multi-objective optimization problems in energy-efficient SWIPT networks may not lead to a satisfactory performance. Therefore, the concept of multi-objective optimization or vector optimization should be adopted for handling conflicting objective functions. Yet again, these factors complicate the design of beamforming for SWIPT.

\subsection{Distributed Transmission Scheme}
In SWIPT systems, there are multiple distributed nodes with different types and tasks, such as information-orient nodes, power-orient nodes, and eavesdropping nodes. In order to design an optimal transmission scheme, it is imperative to collect the knowledge of all the nodes, which results in a significant increment in signalling overhead. Thus, transmission schemes with low signalling overhead and distributed structure are the key to unlock the potential of multiple nodes in SWIPT systems.

\subsection{Adaptive Environment Sensing}
The secrecy performance of an information receiver is affected by its counterparts, including eavesdroppers, power receivers, and other information receivers. For example, eavesdroppers may interfere with the information receiver by sending an artificial noise, so as to intercept more information. Thus, it is imperative to enhance the capability of environment sensing, i.e., sensing the information and behaviours of the eavesdroppers, which is helpful to improve the performance of SWIPT.

\vspace{0.125in}
\section{CONCLUSION}
This article provided a review on SWIPT from both theoretical and technical perspectives. First, we surveyed various SWIPT scenarios, with an emphasis on revealing the challenging issues. Then, we discussed a variety of effective PHY-security techniques, which can effectively improve the performance of SWIPT. In addition, we proposed to use massive MIMO techniques to further enhance SWIPT and showed the performance gain through numerical simulations. Finally, some potential research directions were identified.

\end{spacing}
\end{document}